# Electron microscopy and spectroscopy investigation of atomic, electronic, and phonon structures of NdNiO$_2$/SrTiO$_3$ interface

Yuan Yin[1][‡], Mei Wu[1][‡], Xiang Ding[2][‡], Peiyi He[1][‡], Qize Li[1,3], Xiaowen Zhang[1], Ruixue Zhu[1], Ruilin Mao[1], Xiaoyue Gao[1], Ruochen Shi[1*], Liang Qiao[2*], Peng Gao[1,4,5]*

[1] International Center for Quantum Materials, and Electron Microscopy Laboratory, School of Physics, Peking University, Beijing 100871, China.

[2] School of Physics, University of Electronic Science and Technology of China, Chengdu 611731, China.

[3] Department of Physics, University of California at Berkeley, Berkeley, CA 94720, USA

[4] Interdisciplinary Institute of Light-Element Quantum Materials and Research Center for Light-Element Advanced Materials, Peking University, Beijing 100871, China.

[5] Collaborative Innovation Center of Quantum Matter, Beijing 100871, China.

[‡]These authors contributed equally to the work.

* Corresponding authors. Email: shirc1993@pku.edu.cn, liang.qiao@uestc.edu.cn, pgao@pku.edu.cn.

**Abstract:** The infinite-layer nickelates, proposed as analogs to superconducting cuprates, provide a promising platform for exploring the mechanisms of unconventional superconductivity. However, the superconductivity under atmospheric pressure has only been observed in thin films, indicating the heterointerface is essential. Here, we employed the advanced Scanning Transmission Electron Microscopy-Electron Energy Loss Spectroscopy (STEM-EELS) technique to thoroughly investigate the atomic configuration, layer-resolved electronic states and phonon across the $NdNiO_2/SrTiO_3$ interface. We found the Sr atoms diffusion at the interface, which results in hole doping into oxygen and nickel band to form a p-type interface. A pronounced redshift of the highest-energy optical phonon (HEOP) of $NdNiO_2$ (~78 meV) is observed at the interface, which is primarily attributed to the epitaxial strain. Our work clarifies that the effects of the interface on electron and phonon states are mainly due to elemental intermixing and epitaxial strain, which could lay the foundation for future investigations into superconducting mechanisms of infinite-layer nickelates.

In 1999, it was first predicted that nickelates with the spin S=1/2 Ni$^+$ in planar Ni-O structures could be magnetic insulators, with the doped state Ni$^{2+}$ (S=0) similar to that of Cu$^{3+}$ (S=0) in cuprates after doping with holes [1]. Twenty years later, the first experimental observation of superconductivity in infinite-layer nickelates, with T$_c$ = 9~15 K, triggered widespread interest [2]. On the one hand, infinite-layer nickelates share similarities with cuprates. They both exhibit a 3d$^9$ electron configuration when doped, making the $3d_{x^2-y^2}$ band being the dominant one [3,4]. On the other hand, they also exhibit numerous differences. Cuprates fall under the charge-transfer regime, while infinite-layer nickelates belong to the Mott–Hubbard regime [5], resulting in weaker hybridization between O-2p with Ni-3d orbital than O-2p with Cu-3d orbital [6,7]. Besides, unlike cuprates, nickelates have a multi-band character with a "self-doping" effect [3,8], and they do not have long-range magnetic ordering at low temperatures [9].

These similarities and differences have sparked controversies surrounding the superconductivity pairing mechanism [8,10], long-range antiferromagnetic order [11,12], charge order [13–15], Kondo effect [16], interfacial effects [17–19] of nickelates, and so on. Among these, the interfacial effect has always been a contentious question [17–19] since bulk and powder infinite-layer nickelates have not shown superconductivity to date [17,20]. Calculations have shown that nickelate interfaces of different configurations have different electronic structures [19,21], and the electronic and atomic reconstructions are unavoidable for a polar interface [22]. Experimentally, infinite-layer nickelates with varying thicknesses [23] and different compressive-strain substrates [24] show different superconducting transition temperatures. These computational and experimental results raise questions about whether the interface plays a decisive role in the superconductivity of infinite-layer nickelates. Therefore, it is crucial to clarify the impact of the interface from a microscopic perspective and to reveal its role in superconductivity. Since traditional experimental techniques such as X-ray Absorption Spectroscopy (XAS) and Resonant Inelastic X-ray Spectroscopy (RIXS) can only measure the overall properties without sufficient spatial resolution, and thus cannot recognize the contribution of the interface exclusively. Herein, we employ the advanced scanning transmission electron microscopy-electron energy loss spectroscopy (STEM-EELS) technique with atomic resolution, to probe the electronic and phonon structures at the interface.

In this study, we found an A-site atoms intermixing interface, leading to the formation of stoichiometric $Nd_{1-x}Sr_xNiO_2$ near the interface, which is equivalent to hole doping to form a p-type interface [25]. Our STEM-EELS results show that the holes are partly doped into the 3d orbitals of nickel and partly into the 2p orbitals of oxygen, i.e., with a joint final state of $3d^8$ and $3d^9L$, where L denotes a ligand hole. Furthermore, the highest-energy optical phonon (HEOP) of $NdNiO_2$ (~78 meV) was observed to undergo a pronounced redshift at the interface, primarily attributed to the epitaxial strain according to the calculation results. Thus, we can conclude that the substrate $SrTiO_3$ can provide $NdNiO_2$ films with doping holes through elemental diffusion, similar to the case of $CaCuO_2$/$SrTiO_3$ interfacial superconductivity [26,27]. Furthermore, such interface leads to the softening of the highest-energy optical phonon in $NdNiO_2$. Our findings demonstrate the ability to precisely correlate the atomic structure, electronic states, and phonon structure, thus to understand the underlying mechanism of nickel-based superconductors better.

Infinite-layer nickelates thin films without capping layer were prepared on $SrTiO_3$ (001) substrates by pulsed laser deposition and topotactic reduction (see Supplementary Methods for details). An atomically resolved high-angle annular dark-field (HAADF) image of the interface is shown in Fig. 1(a). To reveal the oxygen configuration near the interface, the annular bright-field (ABF) image is also recorded in Fig. 1(b). The atomically resolved elemental maps of energy dispersive X-ray spectroscopy (EDS) in Fig. 1(c) reveal the arrangements of Nd, Sr, Ni and Ti atoms and confirm that there exists elemental intermixing at the interface. Instead of B-site atoms mixing [19], A-site atoms (Nd and Sr) intermixing was detected, where Sr is the dominant one. The integral line profiles of the corresponding elements are shown in Fig. 1(d), where layer 0 indicated by the arrow is the pronounced mixing layer of Nd and Sr. No notable diffusion of Ti or Ni is observed in Fig. 1(c) and Fig. 1(d). Based on these results, the atomic configuration across the interface is schematically drawn in the inset of Fig. 1(b). Due to the presence of O atoms in the mixed atomic layer, the Ni-O configuration closest to the interface is a pyramidal configuration, different from the Ni-O planar tetragonal configuration of the other layers, resulting in a different coordination environment of the closet layer from others. To delve into the strain near the interface, we performed peak fitting of the A-site atom positions on the HAADF. Then, we calculated the scaled lattice parameter using $SrTiO_3$ far away from the interface as reference, as shown in Fig. 1(e). The in-plane scaled lattice parameter exhibits minor variation, while the out-of-plane one near the interface (within the first two unit cells) shows significant changes. The out-of-plane compressive strain of $NdNiO_2$ near the interface is around 2.6% compared to unstrained bulk $NdNiO_2$.

Here, the observed intermixing layer of Nd and Sr is expected to help to alleviate the "polarization catastrophe", as shown in Supplemental Material, Fig. S1(a) and (b). Therefore, such atomic reconstruction at the interface suppresses potential divergence, thereby excluding the influence of the electric field on the electronic structure [19], which will not be discussed in the following analysis. Meanwhile, this intermixing is equivalent to Sr-doping into $NdNiO_2$, effectively leading to hole doping. As moving away from the interface, the concentration of doped holes decreases due to the reduced concentration of Sr element, resulting in electronic structure changes as expected.

To figure out the changes in the electronic structures near the interface, we acquired the layer-resolved energy-loss near-edge fine structure (ELNES) on the first five layers of $NdNiO_2$ near the interface. The corresponding HAADF image in Fig. 2(a) illustrates the five layers near the acquired interface, with each layer marked by colored dots numbered from 1 to 5, respectively. The O-K edge in Fig. 2(b) exhibits apparent distinctions between different layers. For better comparison, the O-K edge of layer 5 represented by a dashed line was shifted to overlap with other spectral lines. Specifically, the O-K of layer 1 has a distinct pre-peak at about 529 eV, with the peak intensity gradually decreasing when moving away from the interface. Starting from the third layer, the small peak turns into the shoulder and nearly disappears when approaching to layer 5, which is characteristic of bulk $NdNiO_2$, consistent with previous studies [19,28]. Fig. 2(d) shows the additional intensity integrated from 528 to 530 eV relative to layer 5 gradually decreases from layer 1 to layer 4. The pre-peak of the O-K edge reflects the hybridization degree of O-2p and transition metal orbital [29,30]. It is very pronounced in perovskite $NdNiO_3$, where Ni is in a $3d^7L$ ground state in the metallic phase. However, it nearly vanishes in the bulk infinite-layer $NdNiO_2$, where Ni is in a $3d^9$ state [18,28]. Therefore, the emergence of the pre-peak implies the presence of a $3d^9L$ state, which may be induced by the diffusion of Sr elements at the interface, leading to hole doping.

Similarly, the Ni-$L_3$ spectral line of layer 5 was overlaid onto other spectra using dashed lines for better comparison in Fig. 2(c), and the corresponding $L_3$ peak position for each layer was extracted, as shown in Fig. 2(e). The $L_3$ peak exhibits a blue shift when moving closer to the interface, with the peak of layer 1 shifting by nearly 0.5 eV relative to layer 5, indicating an increase in the valence state of Ni and a tendency toward the $3d^8$ state near the interface. It is worth noting that this increase is neither linear nor monotonic, consistent with results reported in previous literature [28]. Additionally, the main peak gradually widens from layer 5 to layer 2 and a high-energy shoulder appears, particularly noticeable in layer 2. This spectral feature is consistent with previous XAS [7] and EELS [28] results reported for $NdNiO_2$ superconducting samples with varying Sr doping, as a sign of hybridization between the O 2p and Ni 3d states. Comparable high-energy shoulder features have previously been reported in the XAS spectra of superconducting cuprates, which have been ascribed to ligand hole state [26,31–33]. The atomic configuration of $Nd_{1-x}Sr_xNiO_{2.5}$ in Fig. 1(b) at the interface further increased the valence state of Ni, resulting in further blueshift of layer 1. The complete $L_{2,3}$ white line of Ni is displayed in Fig. S2(b). To understand the effect of strain on the electronic structure, we performed the DFT calculation shown in Fig. S3. According to the results, the shift in the partial density of states (PDOS) near the Fermi surface with 2.6% strain is negligible, thus the influence of strain has been ruled out here. These results confirm that the diffusion of Sr in the vicinity of the interface is equivalent to hole doping, forming $3d^8$ and $3d^9L$ states. Additionally, the $M_{4,5}$ edges of Nd were also investigated as shown in Fig. S2(c). There are no distinguishable energy shifts and shape changes, indicating that the 4f electronic state of the rare earth element Nd exhibits low sensitivity to such local electron or atom reconstructions [18].

Electron-phonon coupling (EPC) has been considered to contributed to conventional [34] and high-temperature superconductivity [35,36]. Several previous works of DFT calculations claimed that the EPC was too small to contribute to the superconductivity in $NdNiO_2$ [37–40]. However, a recent work using the ab initio GW theory proposed that the EPC contributes significantly to the superconducting temperature of $NdNiO_2$ [41]. Therefore, it is meaningful to study phonons near the interface experimentally. However, since the structure of $NdNiO_2$ has no Raman activity [42], the phonon-related experiments were rarely reported in this system, not to mention the knowledge of phonon spectrum near the interface. To fill this gap, we measured the phonon spectrum near the $NdNiO_2$/$SrTiO_3$ interface (Fig. 3(b)) with the corresponding HAADF for the acquisition region (Fig. 3(a)), where the white dashed line referring to the interface position. The phonons of $NdNiO_2$ bulk (red) and the interface(purple), extracted from the measured EEL spectrum, are shown in Fig. 3(c). The highest-energy optical phonon (HEOP) of $NdNiO_2$ (~78 meV) exhibits an obvious redshift near the interface (from red arrow to purple arrow). To investigate the cause of the phonon changes in $NdNiO_2$, we employed Density Functional Perturbation Theory (DFPT) to calculate the $NdNiO_2$ phonons without and under 2.6% out-of-plane strain. The calculated phonon dispersions and phonon density of states (PDOS) are shown in Fig. S4(a) and S4(b), and their corresponding phonon eigenvectors are presented in Fig. S5. To better compare with the experimental results, we convoluted the calculated PDOS with a Gaussian peak of 12 meV full width at half maximum (FWHM) in Fig. 3(d). The calculated results are in good agreement with the experimental results, showing an evident redshift of the HEOP while other phonon modes remain largely unchanged. Therefore, we conclude that the strain plays a crucial role in the phonon softening near the interface, while the influence of electron-phonon coupling should also not be underestimated as well [41].

$NdNiO_2$ is the parent oxide to $Nd_{1-x}Sr_xNiO_2$, which exhibits superconductivity when x ranges from 10% to 25% [43,44]. Therefore, the doping concentration plays a critical role in the emergence of superconductivity. For NdNiO$_2$ films grown on SrTiO$_3$, our experimental results indicate that the Sr diffuses gradiently into the film, serving as a source of holes. Therefore, the regions with doping concentrations within the range of the superconducting dome can exhibit superconducting properties [45,46]. The STEM-EELS technique is superb for mechanism studies on such an atomic-scale region. Based on this, we hope our study provides valuable insights for further investigation into the superconducting mechanism in infinite-layer nickelates.

In addition, a prior experiment achieved superconductivity with $Tc$=40 K in (CaCuO$_2$)$_m$/(SrTiO$_3$)$_n$ superlattices [26]. Subsequent EELS results demonstrated that at the CaCuO$_2$/SrTiO$_3$ interface, oxygen ions are incorporated into the Ca plane, acting as apical oxygen for Cu and supplying holes to the CuO$_2$ planes, thus facilitating superconductivity [27]. The hole doping is restricted to within 1-2 unit cells near the interface. These experiment results align with our observation of Sr-induced hole doping at the NdNiO$_2$/SrTiO$_3$ interface. Therefore, (NdNiO$_2$)$_m$/(SrTiO$_3$)$_n$ superlattices remain to be fabricated and are well worth further exploring.

As mentioned, hole doping along with epitaxial strain, influences the electronic and phonon structures of the NdNiO$_2$ film at the interface. It was verified that infinite-layer nickelates with different compressive-strain substrates show different superconducting transition temperatures [24]. Several studies on Ruddlesden-Popper phase nickelates have realized 80 K superconducting transition temperature at high pressures [47,48]. However, previous studies primarily focused on the impact of strain on the electronic structure, overlooking its effects on the phonon structure [48]. Our results indicate that 2.6% epitaxial strain has a minimal effect on the electronic structure but contributes to phonon softening at the interface. Therefore, strain engineering could be an effective approach for designing interfaces with specific phononic properties, thereby offering a new direction for exploring potential mechanisms to enhance superconducting performance.

In summary, we examined the p-type NdNiO$_2$/SrTiO$_3$ interface and revealed that elemental intermixing and epitaxial strain at the interface significantly impact the electronic and phonon structures. The diffusion of Sr elements from the substrate, serving as a hole source, introduces gradient doping to the films. Layer-resolved O-K edge and Ni-L edge results show that the doping holes are not only into O 2p but also into Ni 3d orbital, distinguishing them from charge-transfer cuprates and showing multiband characteristics [28]. Additionally, the redshift of the highest-energy optical phonon of the $NdNiO_2$ film at the interface was observed, mainly attributed to strain when comparing the experiment and DFPT calculation results. Our atomic-scale study could be essential for elucidating macroscopic experimental phenomena in infinite-layer nickelates and help to understand the underlying superconductivity mechanisms.

**Acknowledgments**

The work was supported by the National Natural Science Foundation of China (52125307 (to P.G.), 12404192 (to R.C.S), 12274061 (to L. Q.)). L.Q. acknowledges Key Research and Development Program from the Ministry of Science and Technology (2023YFA1406301). P.G. acknowledges the support from the New Cornerstone Science Foundation through the XPLORER PRIZE. We acknowledge Electron Microscopy Laboratory of Peking University for the use of electron microscopes. We acknowledge the High-performance Computing Platform of Peking University for providing computational resources for the DFT and DFPT calculation.

**Figures and captions**

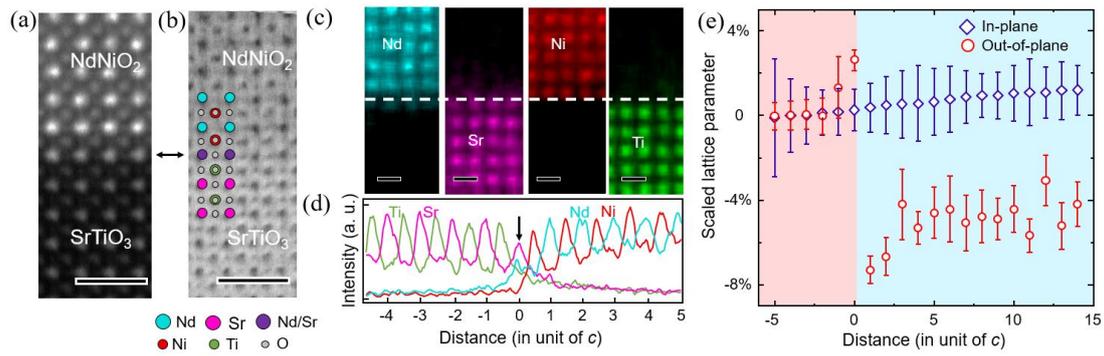

FIG. 1. **The atom arrangement of NdNiO$_2$/SrTiO$_3$ interface.** (a) The HAADF and (b) ABF of NdNiO$_2$ on SrTiO$_3$ substrate, where the elements of the atoms are determined according to the image contrast and energy dispersive X-ray spectroscopy (EDS). The scale bar is 1 nm. (c) The EDS of Nd, Sr, Ni and Ti across the interface with atomic resolution. The scale bar is 5 Å. Intermixing of Nd and Sr(white line) occurs at the layer indicated by the bidirectional arrow layer between (a) and (b). (d) The Sr, Ti, Nd, Ni intensity line profiles of EDS across the interface. The arrow indicts the intermixing layer position. (e) The in-plane and out-of-plane scaled lattice parameters line profile. The error bar is the standard deviation(s.d.).

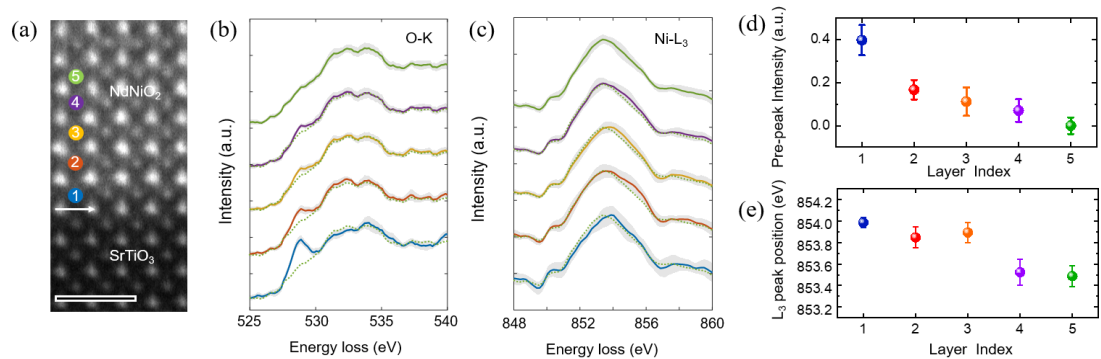

FIG. 2. **Atomically resolved O-K and Ni-L$_3$ edge near the interface.** (a) A STEM-HAADF image of the NdNiO$_2$/SrTiO$_3$ interface, the white arrow indicates the intermixing layer, and the 5 atomic layers away from the interface are marked with numbered colorful dots, referring to the layer 1 to 5 respectively. The scale bar is 1 nm. (b) The O-K edge and (c) Ni-L$_3$ peak of the corresponding five layers, the gray shadow is the standard deviation(s.d.) of the spectral lines and the dotted lines mark the characteristics of the layer 5 for better comparison. (d) The O-K pre-peak intensity integrated over 528 to 530 eV of different atomic layers. (e) The L$_3$ peak position of Ni in different atomic layers. The error bar is the s.d..

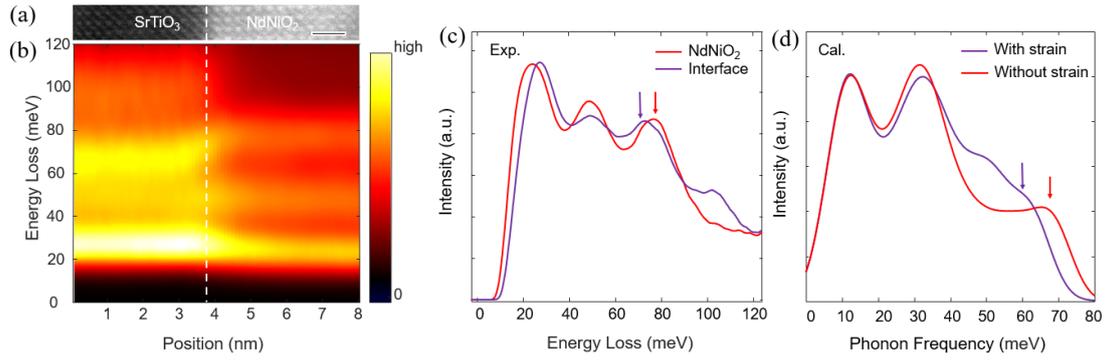

FIG. 3. **The measured and calculated phonon structure across the interface.** (a) The HAADF showing the region $NdNiO_2/SrTiO_3$ for phonon measurement, the scale bar is 1 nm. (b) The phonon mapping of $NdNiO_2/SrTiO_3$, where the interface is labeled by the dashed line. (c) The phonon spectra of $NdNiO_2$ (red), and the interface (purple) extracted from (b), where the small peak (~102 meV) of the interface is contributed from the $SrTiO_3$. (d) The convolution results of calculated PDOS for bulk $NdNiO_2$ without (red) and under 2.6% out-of-plane strain (purple) with a Gaussian peak of 12 meV full width at half maximum (FWHM).

# Electron microscopy and spectroscopy investigation of atomic, electronic, and phonon structures of NdNiO$_2$/SrTiO$_3$ interface

Yuan Yin[1][‡], Mei Wu[1][‡], Xiang Ding[2][‡], Peiyi He[1][‡], Qize Li[1,3], Xiaowen Zhang[1], Ruixue Zhu[1], Ruilin Mao[1], Xiaoyue Gao[1], Ruochen Shi[1*], Liang Qiao[2*], Peng Gao[1,4,5]*

[1] International Center for Quantum Materials, and Electron Microscopy Laboratory, School of Physics, Peking University, Beijing 100871, China.

[2] School of Physics, University of Electronic Science and Technology of China, Chengdu 611731, China.

[3] Department of Physics, University of California at Berkeley, Berkeley, CA 94720, USA

[4] Interdisciplinary Institute of Light-Element Quantum Materials and Research Center for Light-Element Advanced Materials, Peking University, Beijing 100871, China.

[5] Collaborative Innovation Center of Quantum Matter, Beijing 100871, China.

[‡]These authors contributed equally to the work.

\* Corresponding authors. Email: shirc1993@pku.edu.cn, liang.qiao@uestc.edu.cn, pgao@pku.edu.cn.

1. Method

Sample Preparation. Perovskite NdNiO$_3$ thin films (10 nm) were grown on SrTiO$_3$ (001) substrates by pulsed laser deposition using a 248 nm KrF excimer laser. A laser fluence of 1.2 J cm$^{-2}$ was used to ablate the target. The substrate temperature was controlled at 620 °C with an oxygen pressure of 200 mTorr. The thin films were cooled to room temperature under the same oxygen environment after growth. Then, the Perovskite samples were reduced to the infinite-layer phase using topotactic reduction method. The perovskite films were vacuum-sealed (<0.1 mTorr) together with 0.1 g of solid-state CaH$_2$ powder. The temperature profile of the reduction procedure followed a trapezoidal shape. The warming and cooling rates were 10 °C min$^{-1}$. On the plateau, reduction was held at a steady temperature for an optimized time of 2 h. Reduction temperatures of 200 °C were used to produce the NdNiO$_2$ films. Samples had no SrTiO$_3$-capping layers.

Preparation of TEM sample. The TEM specimens were initially thinned by mechanical polishing and subsequently subjected to argon ion milling. The ion milling process was performed using a PIPS™ (Model 691, Gatan, Inc.) with an accelerating voltage of 3.5 kV until a hole appeared. To remove the surface amorphous layer and minimize irradiation damage, low-voltage milling was then conducted at an accelerating voltage of 0.3 kV.

Electron microscopy characterization and image analysis. The HAADF images were recorded at 300 kV using an aberration-corrected FEI Titan Themis G2. The convergence semiangle for imaging was 30 mrad, and the collection semiangle range was 39–200 mrad for HAADF imaging. The ABF image were acquired at 300 kV using aberration-corrected JEM-ARM300F GRAND ARM300. The convergence semiangle for ABF was 24 mrad, and the collection semiangle range was 8.5–37.7 mrad.

Atom positions were determined by simultaneously fitting with two-dimensional Gaussian peaks using a custom-written MATLAB code. The in-plane and out-of-plane scaled lattice parameter in Fig.1(e) are calculated as $(a_{ip} - a_{STO})/a_{STO}$ and $(a_{op} - a_{STO})/a_{STO}$.

EELS data acquisition and processing. The EELS data was acquired on a Nion U-HERMES200 microscope equipped with monochromator and aberration correctors which was operated at 60 kV to reduce the beam damage. The probe convergence semi-angle was 35 mrad and the collection semi-angle was 24.9 mrad. The core-level EEL spectra in Fig. 2(b) and Fig. 2(c) were recorded as 24×60 spectrum image from 1.6 nm × 4 nm region. The dwell time was 250 ms per pixel, and the dispersion was 0.1663 eV per channel. The phonon spectra shown in Fig.3(b) was collected using an off-axis experimental setup. In this case, the electron beam was shifted 60 mrad from the optical axis along the [010] direction of $SrTiO_3$ to significantly minimize the effects of long-range dipole scattering. The data was also recorded over an 8 nm range across the interface with a resolution of 80×10 pixels and an energy dispersion of 0.5 meV/channel with dwell time of 800ms per pixel.

All acquired EEL spectra were processed using the Gatan Microscopy Suite software and custom-written MATLAB toolbox. The background of core-level EEL spectra were fitted and subtracted using the power law function. Then, the block-matching and three dimensional filtering (BM3D) algorithm were applied to remove gaussian noise. Afterwards, the sample drift was corrected by aligning the interface. The spectra were accumulated along the direction parallel to the interface to obtain a line-scan data with good signal-to-noise ratio. For the vibrational spectra, the spectra was multiply by $E^2$ to remove the background effect, where E represents the energy. Lucy–Richardson deconvolution was employed to ameliorate the broadening effect caused by the finite energy resolution, taking the ZLP as the point spread function.

**Ab initio calculations.** Density functional theory calculations were performed using Quantum ESPRESSO [1,2] with the Perdew-Burke-Ernzerhof for solid (PBEsol) [3] exchange-correlation functional and the norm-conserving pseudopotentials (NCPP) available from the PseudoDojo [4]. Before computing structural relaxations, the parameters of $NdNiO_2$ are a=b=3.92 Å, c=3.28 Å for the unstrained one while a=b=3.92 Å, c=3.19 Å for the counterpart under 2.6% out-of-plane strain. The crystal structure is fully relaxed with an energy convergence criterion of $10^{-8}$ eV, force convergence criterion of $10^{-4}$ eV/Å. The energy cutoffs of charge density and wave function are set to be 400 Ry and 100 Ry, respectively. The first Brillouin zone was sampled with the 12 × 12 × 12 k points mesh. Based on DFPT (density functional perturbation theory) [5], the QE package is utilized to calculate the phonon spectra and the phonon DOS of $NdNiO_2$ in DFPT calculation. The 12 × 12 × 12 k grids and 4 × 4 × 4 q grids are carried out the self-consistent field calculation and DFPT calculation, respectively.

**Author contributions**

## 2. Supplemental Figures

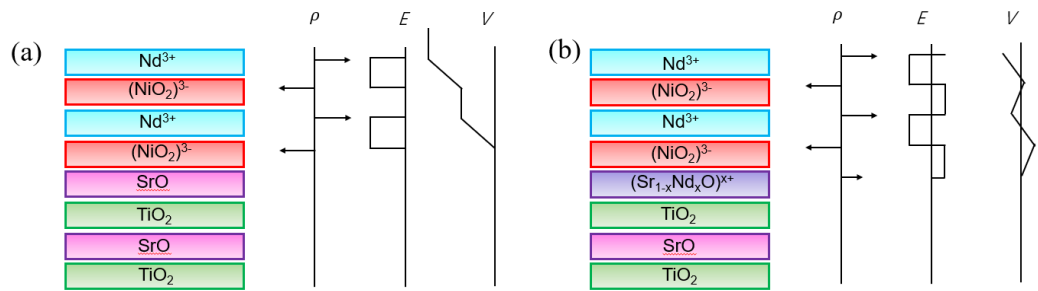

FIG. S1 **Schematic diagrams of the interface polarization.** (a) The unreconstructed and (b) reconstructed p-type interface (SrO terminated plane of SrTiO$_3$), with their $\rho$, electrical field and potential showing different situation.

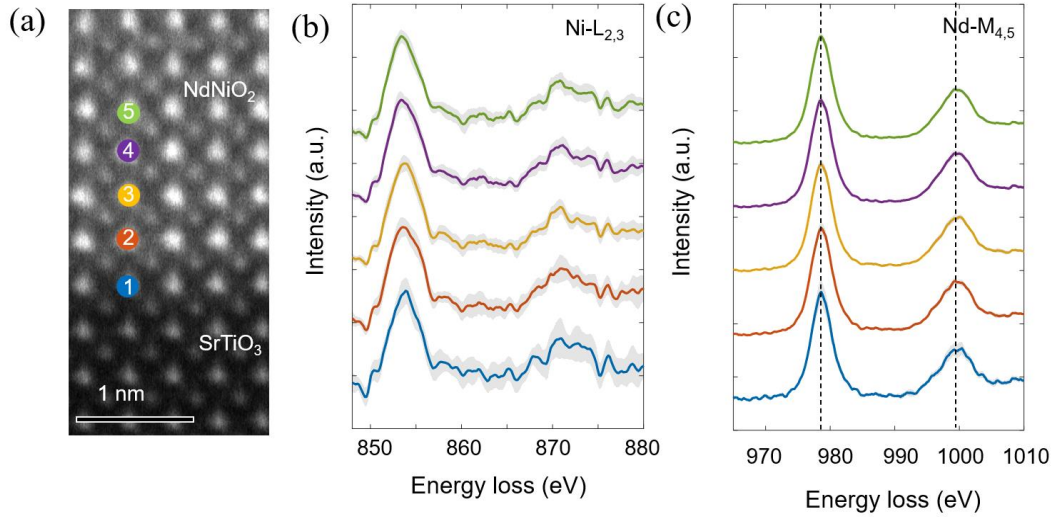

FIG. S2 **Layer-resolved electronic structure information.** (a) The STEM-HAADF picture near the NdNiO$_2$/SrTiO$_3$ interface, the 5 atom layers from the interface up are marked with colored dots with numbers, corresponding to the layer 1 to 5 respectively. (b) The Ni-L$_{2,3}$ and (c) Nd-M$_{4,5}$ edge white lines of 5 layers indicated in (a). The gray shadow is the s.d. of the spectral line. The dashed lines indicate Nd-M$_5$ and M$_4$ peak position respectively, showing no distinguishable shift.

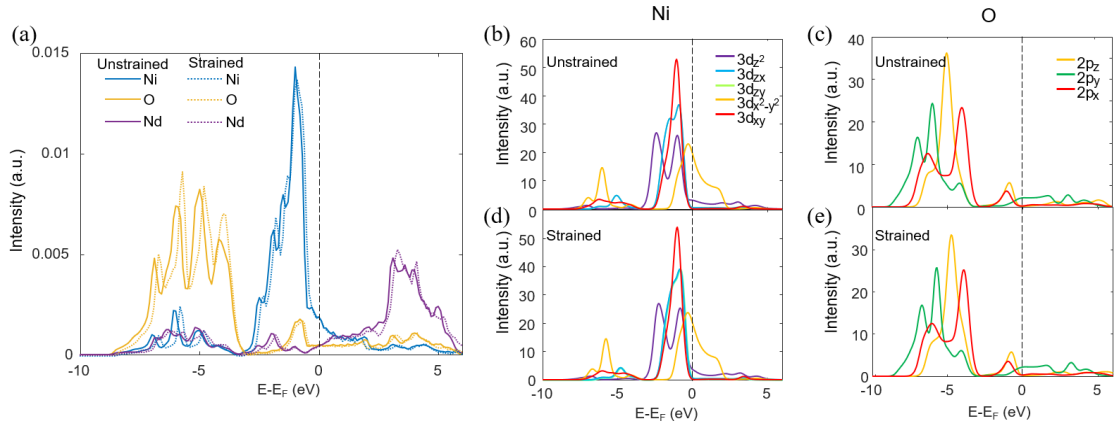

FIG. S3 **The effect of strain on NdNiO$_2$ electronic states by DFT calculations.** (a) The partial density of states (PDOS) of unstrained (solid line) and 2.6% strained (dashed line) NdNiO$_2$ on Ni, O and Nd atoms. The PDOS of unstrained (b, c) and strained (c, e) NdNiO$_2$ on Ni(b, d) and O(c, e) orbitals.

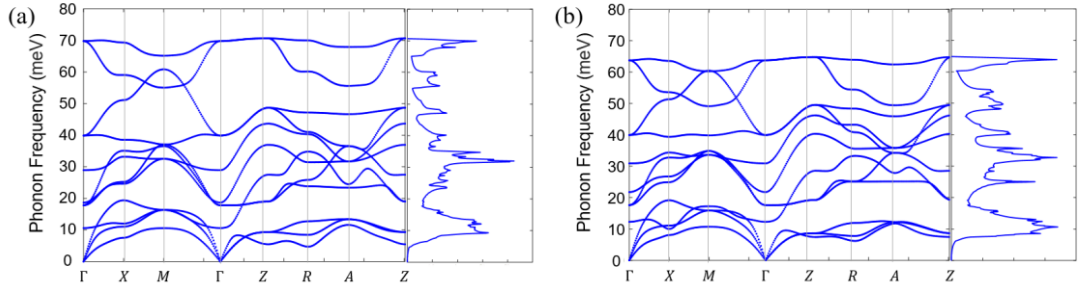

FIG. S4 **The effect of strain on NdNiO$_2$ phonon dispersion and modes by DFT calculations.** DFPT calculations of phonon dispersion and PDOS for bulk NdNiO$_2$ without strain (a) and under 2.6% out-of-plane strain (b).

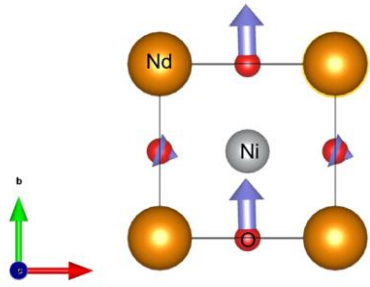

FIG. S5 The eigenvectors of the highest energy LO/TO phonon (~78meV) around Γ point of NdNiO$_2$.